\documentclass[12pt]{article}
\usepackage{amsmath,amssymb,amsfonts,epsfig,graphicx,euscript}

\newcommand{\bse}{\begin{subequations}}
\newcommand{\ese}{\end{subequations}}
\newcommand{\be}{\begin{equation}}
\newcommand{\ee}{\end{equation}}
\newcommand{\bea}{\begin{eqnarray}}
\newcommand{\eea}{\end{eqnarray}}
\newcommand{\ba}{\begin{array}}
\newcommand{\ea}{\end{array}}

\newcommand{\h}{{\frac{1}{2}}}
\newcommand{\ie}{{\it i.e.}}

\begin{document}
\hfill%
\vbox{
    \halign{#\hfil        \cr
           IPM/P-2010/014\cr
                     }
      }
\vspace{1cm}
\begin{center}
{ \Large{\textbf{A D2-brane in the Penrose limits of AdS$_4\times$CP$^3$}
\\}}

\vspace{2cm}
{M. Ali-Akbari}\\
{School of Physics,\\ Institute for Research in Fundamental Sciences (IPM),\\
P. O. Box 19395-5531, Tehran, Iran }\\
{\texttt{E-mail:aliakbari@theory.ipm.ac.ir}}
\end{center}

\vspace{1cm}
\begin{center}
\textbf{Abstract}
\end{center}
We consider a D2-brane in the pp-wave backgrounds obtained from
AdS$_4\times$CP$^3$ when electric and magnetic fields have been
turned on. Upon fixing the light-cone gauge, light-cone Hamiltonian
and BPS configurations are obtained. In particular we study BPS
configurations with electric dipole on the two sphere giant and a
giant graviton rotating in transverse directions. Moreover we show
that the gauge field living on the D2-brane is replaced by a scalar
field in the light-cone Hamiltonian. We also present a matrix model
by regularizing (quantizing) 2-brane theory.

\newpage


\newpage
\section{Introduction}
AdS$_5$/CFT$_4$ correspondence identifies ${\cal{N}}=4$ SU(N)
superconformal gauge theory to type IIB superstring theory on the
maximally supersymmetric AdS$_5\times$S$^5$ background. This
correspondence is a weak/strong coupling corresponding and this
makes it a powerful tool to compute the strong coupling region of
either theory using the weak coupling of the other. Albeit helpful,
this property makes difficult to test AdS/CFT duality explicitly
since neither type IIB superstring on AdS$_5\times$S$^5$ background
nor strong coupling gauge theory are well-understood.

Another maximal supersymmetric solution of type IIb supergravity is
pp-wave and it can be obtained by taking Penrose limit of
AdS$_5\times$S$^5$. The superstring theory on this background was
explicitly solved \cite{Metsaev}. Therefore in the pp-wave
background we know the string spectrum and can check that whether
the same spectrum exists on the gauge theory side. Then we first
need to understand how this specific limit translates to the gauge
theory side. It was argued that the Penrose limit corresponds to
considering a certain section of operators namely BMN operators
\cite{BMN}. Study of AdS/CFT correspondence in this specific limit
opens a new way to test this conjecture more precisely.

Another example of AdS/CFT duality is AdS$_4$/CFT$_3$. ${\cal{N}}=8$
CFT$_3$ was an open question for years and it was finally written in
\cite{Bagger}, the so-called BLG theory which is a ${\cal{N}}=8$
three dimensional superconformal Chern-Simon theory. AdS$_4$/CFT$_3$
tells us that this theory is a suitable candidate to describe
multiple M2-branes. But after a while it was shown that the BLG
theory describes two coincident M2-branes \cite{VanRaamsdonk}. Based
on the BLG model, ABJM theory has been nominated to describe low
energy of multiple M2-branes and to be dual to M-theory on
AdS$_4\times S^7/Z_k$ \cite{ABJM}. The ABJM model is a ${\cal{N}}=6$
three dimensional superconformal $U(N)\times U(N)$ Chern-Simon
theory of level $k$ and $-k$. The duality between ABJM model and
type IIA string theory on AdS$_4\times$CP$^3$ has been found when
$N^{1/5}\ll k\ll N$ \cite{ABJM}.

Pp-wave background has been also studied in the AdS$_4$/CFT$_3$
context. A Penrose limit of AdS$_4\times$CP$^3$ with zero space-like
isometry was obtained in \cite{Nishioka} and string spectrum and
BMN-like operators were obtained. Also, pp-wave metrics with one
flat direction and two space-like isometries were found in
\cite{Grignani,Grignani2}. In this paper we consider a D2-brane in a
general pp-wave background \cite{Grignani} and will then find
light-cone (LC) Hamiltonian and BPS configurations with electric
field. This paper is organized as follows. In the next section we
will review pp-wave backgrounds and in section 3 the LC Hamiltonian
for a D2-brane in pp-wave backgrounds will be obtained by using LC
gauge. Then we replace the gauge field on the D2-brane by a scalar
field and find a matrix model by applying a suitable prescription.
In section 4, BPS configurations are given. The last section is
devoted to discussion.

\section{Pp-wave backgrounds}
In this section we will review three pp-wave backgrounds which are
coming from AdS$_4\times$CP$^3$ by taking Penrose limit. One of the
differences between them is concerned with the number of space-like
isometries. A general form of these metrics has been written in
\cite{Grignani} which leads to three pp-wave backgrounds by choosing
appropriate parameters. It is important to notice that the only
meaningful pp-wave backgrounds in AdS/CFT context are those which
are derived from Penrose limit of AdS$_4\times$CP$^3$. The general
form
of pp-wave geometry is given by %
\be\begin{split} %
 ds^2&=-4dx^+dx^-+\sum_{\hat{i}=1}^4\Big(du_{\hat{i}}^2-u_{\hat{i}}^2(dx^+)^2\Big)
 +\sum_{a=1}^2\bigg[dx_a^2+dy_a^2  \cr %
 &+(\xi^2_a-\frac{1}{4})(x_a^2+y_a^2)(dx^+)^2+2\bigg((\xi_a-2C_a)x_ady_a-(\xi_a+2C_a)y_adx_a\bigg)dx^+\bigg],
\end{split}\ee %
and by the following parameters we have %
\bse\begin{align} %
 \label{pp1}\rm{no\ flat\ direction}&\leftrightarrow\xi_a=C_a=0,\\
 \rm{one\ flat\ direction}&\leftrightarrow\xi_1=\frac{1}{2},\xi_2=b+\frac{1}{2},C_1=\frac{1}{4},C_2=0, \\
 \rm{two\ flat\ directions}&\leftrightarrow\xi_a=-\frac{1}{2},C_a=\frac{1}{4},
\end{align}\ese %
where $b$ is an arbitrary parameter. In addition in
AdS$_4\times$CP$^3$ background there are two- and four-form
RR fields which after taking Penrose limit become %
\be\begin{split} %
 C_{+ij}=-\frac{1}{g_s}\epsilon_{ijk}u_k,\ \ C_+=-\frac{1}{g_s}u_4,
\end{split}\ee %
where $i,j=1,2,3$ and $g_s$ is IIA string coupling constant.

AdS$_4\times$S$^7$ is a maximally supersymmetric background. After
taking the $Z_k$ orbifolding of S$^7$ and reducing the M-theory
background AdS$_4\times$S$^7/Z_k$ to type IIA string background
AdS$_4\times$CP$^3$, 24 out of 32 killing spinors remain
\cite{Nishioka1}. It was shown that the case \eqref{pp1} also
preserves 24 supercharges \cite{Sugiyama}. More supersymmetric
pp-waves in M-theory and their dimensional reduction to D0-brane or
pp-waves in type IIA and T-dualisation to solutions in type IIB
theory are studied in \cite{pope}. Moreover, in each case of the
above pp-wave backgrounds coming from AdS$_4\times$CP$^3$, the
minimum bosonic symmetry is a $SO(3)$ rotation acting on $u^i$ as
well as the translation symmetry in $x^+$ and $x^-$ directions.

\section{Light-cone Hamiltonian}
The low energy effective action for a D2-brane in general form of pp-wave background
is %
\be\begin{split}\label{fullaction} %
 S&=\int d\tau d^2\sigma\sqrt{-\det N}+\int C^{(3)} +\int C^{(1)}\wedge F\cr
  &=\int d\tau d^2\sigma{\cal{L}},
\end{split}\ee %
where  %
\bse\begin{align}
 \label{metric}g_{\hat{\mu}\hat{\nu}}=&-4\partial_{\hat{\mu}}x^-\partial_{\hat{\nu}}x^+
 +\big[(\xi^2_a-\frac{1}{4})(x_a^2+y_a^2)-u_{\hat{i}}^2\big]\partial_{\hat{\mu}}x^+\partial_{\hat{\nu}}x^+
 +\partial_{\hat{\mu}}x^I\partial_{\hat{\nu}}x^I\\ \nonumber
 &+2(\xi_a-2C_a)x_a\partial_{\hat{\mu}}y_a\partial_{\hat{\nu}}x^+-2(\xi_a+2C_a)y_a\partial_{\hat{\mu}}x_a\partial_{\hat{\nu}}x^+,\\
 F_{\hat{\mu}\hat{\nu}}=&\partial_{\hat{\mu}}A_{\hat{\nu}}-\partial_{\hat{\nu}}A_{\hat{\mu}},\\
 N_{\hat{\mu}\hat{\nu}}=&g_{\hat{\mu}\hat{\nu}}+F_{\hat{\mu}\hat{\nu}},
\end{align}\ese %
and $x^I=\{u^{\hat{i}},x^a,y^a\}$. In the LC gauge we fix a part of
the area preserving diffeomorphism invariance which mix world-volume
time and spatial coordinates. In order to fix LC gauge we separate
the space and time indices on the brane world-volume as
$\sigma^{\hat{\mu}}=(\tau=\sigma^0,\sigma^r)$, $r=1,2$.
The LC gauge is fixed by choosing  %
\be %
 x^+=\tau.
\ee %
In order to ensure that the above condition is respected by dynamics
we use the time-space mixing part of area preserving diffeomorphism
and set \cite{AliAkbari}%
\be %
 N^{0r}+N^{r0}\equiv G^{0r}=G_{0r}=(g-FgF)_{0r}=0.
\ee %
$G^{\hat{\mu}\hat{\nu}}$ is the symmetric part of
$N^{\hat{\mu}\hat{\nu}}$ which has the interpretation of open string
metric \cite{Seiberg} with $G_{\hat{\mu}\hat{\nu}}$ its inverse.

We noted that in the pp-wave background, $x^+$ and $x^-$ are cyclic variables and their conjugate
momenta are constants of motion. Then %
\be\label{p+} %
 p^+=\frac{\partial{\cal{L}}}{\partial(\partial_\tau x^-)}=-\frac{2}{g_s}\sqrt{-\det N}N^{00},
\ee %
and LC Hamiltonian is %
\be %
 {\cal{H}}_{lc}=p^-=\frac{\partial{\cal{L}}}{\partial(\partial_\tau x^+)}.
\ee %
From the above equation we have %
\be\begin{split}\label{hamiltonian} %
 {\cal{H}}_{lc}=p^+\bigg(\partial_\tau x^--&\h\big[(\xi_a^2-\frac{1}{4})(x_a^2+y_a^2)-u_{\hat{i}}^2\big]\cr
 -&\h(\xi_a-2C_a)x_a\dot{y}_a+\h(\xi_a+2C_a)y_a\dot{x}_a\cr
 -&\frac{1}{2p^+g_s}\epsilon^{ijk}u^i\{u^j,u^k\}-\frac{2Bu_4}{3p^+g_s}\bigg).
\end{split}\ee %
Next we should eliminate $\partial_\tau x^-$. Using \eqref{metric}, $N_{00}$ is
\be\begin{split}\label{N00} %
 N_{00}=-4\partial_\tau x^-+(\xi_a^2-\frac{1}{4})(x_a^2+y_a^2)-(u^{\hat{i}})^2+(\dot{x}^I)^2
 &+2(\xi_a-2C_a)x_a\dot{y}_a\cr
 &-2(\xi_a+2C_a)y_a\dot{x}_a.
\end{split}\ee %
Let us recall the definition of $\det N$ which is  %
\be %
 \det N=\det(N_{rs})(N_{00}-N_{0r}N^{rs}N_{s0}),
\ee %
where $N^{rp}N_{ps}=\delta^r_s$. It is important to note that $N^{00}\neq\frac{1}{N_{00}}$ because of off-diagonal electric-magnetic fields
and hence %
\be %
 N^{00}=\frac{\det(N_{rs})}{\det N}.
\ee %
The above two equations together with \eqref{p+} lead to %
\be\label{max1} %
 N_{00}=-(\frac{2}{p^+g_s})^2\det(N_{rs})+N_{0r}N^{rs}N_{s0}.
\ee %
By means of \eqref{max1} the LC Hamiltonian \eqref{hamiltonian} becomes %
\be\begin{split}\label{hamiltonian2} %
 {\cal{H}}_{lc}&=p^+\bigg((\frac{1}{p^+g_s})^2\det N_{rs}-\frac{1}{4}N_{0r}N^{rs}N_{s0}
 +\frac{1}{4}(\dot{x}^I)^2\cr
 &-\frac{1}{4}\big[(\xi_a^2-\frac{1}{4})(x_a^2+y_a^2)-(u^{\hat{i}})^2\big]
 -\frac{1}{2p^+g_s}\epsilon^{ijk}u^i\{u^j,u^k\}-\frac{2Bu_4}{3p^+g_s}\bigg),%
\end{split}\ee %
where Chern-Simon terms has been added. In the case of D2-brane the
first term in the
Hamiltonian is  %
\be\begin{split} %
 \det N_{rs}&=\det g_{rs}+ \det F_{rs} \cr%
 &=\h\{x^I,x^J\}^2+B^2,
\end{split}\ee %
where $B=F_{12}$ and $\{F,G\}=\epsilon^{rs}\partial_r F\partial_s
G$. The second term of \eqref{hamiltonian2} can be simplified by
using the momentum conjugate to the gauge field
which is %
\be %
 p^r_E=\frac{\partial{\cal{L}}}{\partial F_{0r}}=\frac{1}{2g_s}\sqrt{-\det N}N^{0r},
\ee %
and one can then show  %
\be\begin{split} %
 -p^+N_{0r}N^{rs}N_{s0}&=\frac{16}{p^+}p^r_Eg_{rs}p^s_E \cr
 &=\frac{16}{p^+}p^r_E\partial_rX^Ip^s_E\partial_sX^I=\frac{(4p_E^I)^2}{p^+}.
\end{split}\ee %
Putting all these together we find the LC Hamiltonian to be %
\be\begin{split}\label{fhamiltonian} %
 {\cal{H}}_{lc}&=\frac{(2p^I_E)^2}{p^+}+\frac{(p^{\hat{i}})^2}{p^+}+\frac{p^+}{4}\bigg(\frac{2p^a_x}{p^+}
 -(\xi_a+2C_a)y_a\bigg)^2\cr
 &+\frac{p^+}{4}\bigg(\frac{2p^a_y}{p^+}+(\xi_a-2C_a)x_a\bigg)^2
 +\frac{1}{2p^+g_s^2}\{x^I,x^J\}^2+\frac{B^2}{p^+g_s^2}\cr
 &-\frac{p^+}{4}\big[(\xi_a^2-\frac{1}{4})(x_a^2+y_a^2)-(u^{\hat{i}})^2\big]
 -\frac{1}{2g_s}\epsilon^{ijk}u^i\{u^j,u^k\}-\frac{2Bu_4}{3g_s},%
\end{split}\ee %
where the third term in \eqref{hamiltonian2} was
replaced by the following conjugate momenta %
\be\begin{split} %
 p^{\hat{i}}&=\frac{\partial{\cal{L}}}{\partial(\partial_\tau
 u^{\hat{i}})}=-\frac{p^+}{2}\dot{u}_{\hat{i}},\cr
 p^a_x&=\frac{\partial{\cal{L}}}{\partial(\partial_\tau
 x^a)}=-\frac{p^+}{2}\big[\dot{x}_a-(\xi_a+2C_a)y_a\big],\cr
 p^a_y&=\frac{\partial{\cal{L}}}{\partial(\partial_\tau y^a)}=-\frac{p^+}{2}\big[\dot{y}_a+(\xi_a-2C_a)x_a\big].
\end{split}\ee %

\subsection*{Matrix model}%
BMN (BFSS) matrix model is an interesting candidate for DLCQ of
M-theory in terms of D0-branes in maximally supersymetric eleven
dimensional pp-wave background (flat space) \cite{BMN,BFSS}. The
Hamiltonian of this model is obtained as a regularized version of
M2-brane LC hamiltonian in eleven dimensional pp-wave background
\cite{Dasgupta,hoppe}. Moreover another matrix model describing DLCQ
of type IIB string theory on the maximally supersymetric ten
dimensional pp-wave background has been introduced in
\cite{SheikhJabbari}, namely TGMT (Tiny Graviton Matrix Model). By
regularizing spherical D3-brane in the ten dimensional pp-wave
background, the Hamiltonian of the TGMT matrix model is obtained. In
the following, the gauge field on the D2-brane is replaced by a
scalar field and, by using the logic of \cite{Dasgupta,hoppe}, a
matrix model is introduced.

The gauge field living on a D2-brane has only one physical degree of
freedom and it can be replaced by a scalar field in three
dimensions. We are going to replace electric and magnetic fields in
the LC hamiltonian by derivative of scalar field.
In the case of D2-brane, it is easy to show that %
\be\begin{split}\label{DBIaction} %
 {\cal{L}}_{DBI}&=\sqrt{\det g\left(1+\h F_{\hat{\mu}\hat{\nu}}F^{\hat{\mu}\hat{\nu}}\right)}\cr
 &=-\frac{1}{2p}\det g+\frac{p}{2}\left(1+\h F_{\hat{\mu}\hat{\nu}}F^{\hat{\mu}\hat{\nu}}\right),
\end{split}\ee %
where $p$ is a Lagrangian multiplier. Let us define %
\be\label{max2} %
 F^{\hat{\mu}\hat{\nu}}=\beta\epsilon^{\hat{\mu}\hat{\nu}\hat{\alpha}}t_{\hat{\alpha}},
\ee %
where $\beta$ and $t_{\hat{\alpha}}$ are arbitrary constant and
vector respectively. By using the equation of motion for the gauge
field coming from \eqref{DBIaction} together with \eqref{max2} we find %
\be\begin{split} %
 t_{\hat{\alpha}}&=\partial_{\hat{\alpha}}\varphi,\cr
 P^r_E&=\beta\epsilon^{rs}\partial_s\varphi,\cr
 B&=\beta\dot{\varphi}.
\end{split}\ee %
Terms including electric and magnetic fields in LC
Hamiltonian are thus simplified as follows %
\be\begin{split} %
 \frac{(2p^I_E)^2}{p^+}&=\frac{1}{2p^+g_s^2}\{x^I,\varphi\}^2,\cr
 \frac{B^2}{p^+g_s^2}-\frac{2Bu_4}{3g_s}+\frac{p^+}{4}u^2_4&=
 p^+\big(\frac{p_\varphi}{p^+}-\frac{1}{3}u_4\big)^2
 +\frac{5p^+}{36}u_4^2,
\end{split}\ee %
where $\beta=\frac{1}{2\sqrt{2}g_s}$ and
$p_\varphi=\frac{\dot{\varphi}}{\beta g_s}$. Since we are looking
for DLCQ description
we need to compactify $x^-$ on a circle of radius $R_-$ %
\be %
 x^-\equiv x^-+2\pi R_- .%
\ee %
This leads to the quantization of the LC momentum $p^+$ %
\be %
 p^+=\frac{J}{R_-}.
\ee %
By following \cite{hoppe,SheikhJabbari}, we replace $x^{\hat{I}},p^{\hat{I}}$ with
$J\times J$ matrices, \ie
\be\begin{split} %
 x^{\hat{I}}&\leftrightarrow X^{\hat{I}},\cr %
 p^{\hat{I}}&\leftrightarrow JP^{\hat{I}}, %
\end{split}\ee %
together with %
\be\begin{split} %
 p^+\int d^2\sigma&\leftrightarrow \frac{1}{R_-} \rm{Tr},\cr
 \{F,G\}&\leftrightarrow J[F,G],
\end{split}\ee %
where $x^{\hat{I}}=(x^I,\varphi)$. Equation \eqref{fhamiltonian}
then becomes%
\be\begin{split}%
 {\rm{H}}&=R_-{\rm{Tr}}\Bigg[(P^i)^2+\frac{1}{4}\bigg(2P^a_x
 -\frac{1}{R_-}(\xi_a+2C_a)Y_a\bigg)^2+\big(P_\varphi-\frac{1}{3R_-}U_4\big)^2\cr
 &+\frac{1}{4}\bigg(2P^a_y+\frac{1}{R_-}(\xi_a-2C_a)X_a\bigg)^2
 +\frac{1}{2g_s^2}[X^{\hat{I}},X^{\hat{J}}]^2+\frac{5}{36R_-^2}U_4^2\cr
 &-\frac{1}{4R_-^2}\big[(\xi_a^2-\frac{1}{4})(X_a^2+Y_a^2)-(U^i)^2\big]
 -\frac{1}{2R_-g_s}\epsilon^{ijk}U^i[U^j,U^k]\Bigg].%
\end{split}\ee %
Inspired by \cite{hoppe,SheikhJabbari}, this matrix model describes
DLCQ of M-theory on the uplifted pp-wave backgrounds obtained from
AdS$_4\times$CP$^3$.

\section{BPS configuration}
In this section we study BPS configurations involving
electromagnetic fields. The case of our interest is the static
electromagnetic fields. Our solutions include giant graviton and deformed giant
graviton. Moreover a giant graviton rotating in transverse
directions will be found as a BPS state.

\subsection*{Giant-like solution }
We start with the case where $u^i\neq0$ while other fields set to be zero. In this case the LC hamiltonian is %
\be\begin{split} %
 {\cal{H}}_{lc}&=\frac{p^+}{4}\Big(u_i^2+\frac{2}{(p^+g_s)^2}\{u^i,u^j\}^2
 -\frac{2}{p^+g_s}\epsilon_{ijk}u^i\{u^j,u^k\}\Big)\cr
 &=\frac{p^+}{4}\Big(u^i-\frac{1}{p^+g_s}\epsilon^{ijk}\{u^j,u^k\}\Big)^2.
\end{split}\ee %
We consider the following ansatz
\be\label{anzats} %
 u^i=\frac{\alpha}{2}p^+g_sJ^i,
\ee %
where $\alpha$ is a constant and $J^i$'s satisfy %
\be %
 \{J^i,J^j\}=\epsilon^{ijk}J^k,
\ee %
which specifies a two-sphere whose radius is one. By substituting
\eqref{anzats} in the LC Hamiltonian we then have %
\be\begin{split} %
 {\cal{H}}_{lc}&=\frac{1}{16}(p^+)^3g_s^2\alpha^2(1-\alpha)^2
\end{split}\ee %
The usual BPS argument tells us that ${\cal{H}}_{lc}$ is minimized
when %
\be\label{max5} %
 \alpha=0\ \ {\rm{or}}\ \ \alpha=1
\ee %
The above solutions \eqref{max5} are graviton ($\alpha=0$) and giant
graviton \footnote{Giant graviton in the AdS$_4\times$CP$^3$
background is discussed in \cite{Chandrasekhar}.} ($\alpha=1$) where
their radii are zero and $\h p^+g_s$ respectively. These are $\h$BPS
(12 out of 24 in the case \eqref{pp1}) configurations whose LC
energy is zero and preserve $SO(3)$ symmetry.

One can turn on a constant magnetic field on the spherical D2-brane
\ie\ $B=F_{12}=\rm{constant}$. This magnetic field doesn't change
the spherical shape and the radius of giant graviton but moves its
center of mass from $u_4=0$ to $u_4=\frac{3B}{p^+g_s}$.

\subsection*{BIGGons solution}
For the pure electric field, \eqref{fhamiltonian} simplifies to %
\be\begin{split} %
 {\cal{H}}_{lc}&=\frac{4}{(p^+)^2}\bigg((P^i_E)^2+(\tilde{u}^i)^2\bigg)\cr
 &=\frac{4}{(p^+)^2}\left(\big(\tilde{u}^i\pm
 R^{ij}P^j_E\big)^2\mp2\tilde{u}^iR^{ij}P^j_E\right),
\end{split}\ee %
where
$\tilde{u}^i=\frac{(p^+)^{3/2}}{4}\big(u^i-\frac{1}{p^+g_s}\epsilon^{ijk}\{u^j,u^k\}\big)$
and $R^{ij}$ is a $SO(3)$ rotation. Hence, the BPS equation is %
\be\label{max3} %
 \tilde{u}^i=R^{ij}P^j_E.
\ee %
This BPS equation was discussed in section 3.1 of \cite{AliAkbari}
for the case of a three sphere giant graviton where the electric
field is turned on. There, the shape deformation induced by the
electric field sourced by two equally and opposite point charges
placed on the North and South poles of the three spherical brane was
obtained. The findings of \cite{AliAkbari} generalize BIons
\cite{Callan} to spherical D3-brane BIGGons. Remarkably,
\eqref{max3} and its solutions are the same to those found in
\cite{AliAkbari}. In other words we have found BIGGons solutions for
spherical D2-branes. Physically these family of solutions describe
open strings ending on two sphere giant graviton.

\subsection*{Rotating giant graviton solution}
Another family of solutions that we consider are rotating giant
gravitons. We turn on $u^i(\tau,\sigma^r)$, $x^a(\tau)$ and
$y^a(\tau)$ fields and the LC Hamiltonian thus becomes %
\be\begin{split}\label{rotatinggiant} %
 {\cal{H}}_{lc}=&\frac{p^+}{4}\left(u^i-\frac{1}{p^+g_s}\epsilon^{ijk}\{u^j,u^k\}\right)^2+
 \left(p_x^a\pm\frac{1}{2}p^+\alpha_+y_a\right)^2+
 \left(p_y^a\mp\frac{1}{2}p^+\alpha_-x_a\right)^2\cr
 +&\left(\mp\alpha_+-\xi_a-2C_a\right)p_x^ay_a+\left(\xi_a-2C_a\pm\alpha_-\right)p_y^ax_a,
\end{split}\ee %
where %
\be %
 \alpha_\pm^2=(\xi_a\pm2C_a)^2-(\xi_a^2-\frac{1}{4}).
\ee %
If the coefficients of the last two terms in \eqref{rotatinggiant}
are equal they will show an angular momentum. Let us consider case
\eqref{pp1}. In this case $\alpha_\pm=\frac{1}{2}$ and
hence\footnote{Similar solution exists for the
case $\alpha_\pm=-\frac{1}{2}$.} %
\be\begin{split} %
 {\cal{H}}_{lc}=&\frac{p^+}{4}\left(u^i-\frac{1}{p^+g_s}\epsilon^{ijk}\{u^j,u^k\}\right)^2+
 \left(p_x^a\pm\frac{1}{4}p^+y_a\right)^2\cr
 +&\left(p_y^a\mp\frac{1}{4}p^+x_a\right)^2
 \mp\frac{1}{2}(p_x^ay_a-p_y^ax_a).
\end{split}\ee %
The BPS equations are given by %
\be\begin{split}\label{max4} %
 u^i=&\frac{1}{p^+g_s}\epsilon^{ijk}\{u^j,u^k\},\cr
 p_x^a=&\pm\frac{1}{4}p^+y_a,\cr
 p_y^a=&\mp\frac{1}{4}p^+x_a,
\end{split}\ee %
and the LC Hamiltonian is %
\be %
 {\cal{H}}_{lc}=\frac{1}{16}p^+(x_a^2+y_a^2)=L_{x_ay_a}.
\ee %
The above solution \eqref{max4} describes a giant graviton rotating
in $x^a-y^a$ plane whose angular momentum is
$\frac{1}{16}p^+(x_a^2+y_a^2)$. This configuration is
$\frac{1}{4}$BPS and preserves $SO(3)\times U(1)\times U(1)$.
Obviously one can also consider a giant graviton rotating in
$x_1-x_2$ or $y_1-y_2$ plane.

\section{Conclusion}
There are three different pp-wave backgrounds coming from
AdS$_4\times$CP$^3$ where they have a different number of space-like
isometry. We consider a D2-brane in these pp-wave backgrounds and
the LC hamiltonian of this system is found by applying LC gauge
fixing. There is a contribution coming from the gauge field living
on the D2-brane in the LC Hamiltonian considered as a electric and
magnetic fields. We show that in three dimensions these fields are
replaced by derivative of a scalar field. Using the idea of matrix
model \cite{Dasgupta,hoppe}, we propose a matrix theory describing
M-theory on the uplifted pp-wave backgrounds.

We then find BPS configurations. Half-BPS solutions are graviton and
giant graviton with $SO(3)$ symmetry. For pure electric field, we
reproduce BIGGons configurations describing open strings ending on
giant graviton. These are $\frac{1}{4}$BPS configurations.

A giant graviton rotating in transverse directions is another
$\frac{1}{4}$BPS configuration. Our solution has $SO(3)\times
U(1)\times U(1)$ symmetry and rotates in $x^a-y^a$ plane. Rotation
can be easily extended to other planes in transverse directions.

\section*{Acknowledgment}
It is a great pleasure to thank M. M. Sheikh-Jabbari for valuable
discussions and comments. We would like to thank M. Vincon for
reading the manuscript carefully.

\end{document}